\documentclass[aps,pra,twocolumn,a4paper,superscriptaddress,floatfix,10pt,longbibliography]{revtex4-2}

\usepackage{graphicx}
\usepackage{dcolumn}
\usepackage{bm}
\usepackage{color}
\usepackage{amsmath}
\usepackage{amssymb}
\usepackage{hyperref}

\begin{document}

\title{Quantum Monte Carlo simulations of two-dimensional repulsive Fermi gases with population imbalance}

\author{S. Pilati}
\affiliation{School of Science and Technology, Physics Division, Universit{\`a} di Camerino, 62032 Camerino (MC), Italy}
\author{G. Orso}
\affiliation{Universit\' e de Paris, Laboratoire Mat\' eriaux et Ph\' enom\`enes Quantiques, CNRS, F-75013, Paris, France}
\author{G. Bertaina}
\affiliation{Istituto Nazionale di Ricerca Metrologica, Strada delle Cacce 91, 10135 Torino, Italy}

\begin{abstract}
The ground-state properties of two-component repulsive Fermi gases in two dimensions are investigated by means of fixed-node diffusion Monte Carlo simulations.
The energy per particle is determined as a function of the intercomponent interaction strength and of the population imbalance. 
The regime of universality in terms of the s-wave scattering length is identified by comparing results for hard-disk and for soft-disk potentials.
In the large imbalance regime, the equation of state turns out to be well described by a Landau-Pomeranchuk functional for two-dimensional polarons. To fully characterize this expansion, we determine the polarons' effective mass and their coupling parameter, complementing previous studies on their chemical potential.
Furthermore, we extract the magnetic susceptibility from low-imbalance data, finding only small deviations from the mean-field prediction.
While the mean-field theory predicts a direct transition from a paramagnetic to a fully ferromagnetic phase, our diffusion Monte Carlo results suggest that the partially ferromagnetic phase is stable in a narrow interval of the interaction parameter. 
This finding calls for further analyses on the effects due to the fixed-node constraint.
\end{abstract}

\maketitle

\section{Introduction}

In recent years, two-component atomic Fermi gases with imbalanced populations have been the focus of intense theoretical and experimental research activities~\cite{RevModPhys.80.1215}.
On the attractive branch of Feshbach resonances, researchers analyzed the phase separation between superfluid and normal phases beyond the critical polarization corresponding to the Chandrasekhar-Clogston limit~\cite{PhysRevLett.95.230403,PhysRevLett.97.200403,PhysRevLett.100.030401,PhysRevA.79.013616,shin2008phase,PhysRevLett.103.170402,PhysRevLett.102.230402,PhysRevLett.101.070404,PhysRevA.92.063616,PhysRevLett.106.215303}.
On the other hand, the repulsive branch allowed exploring the Stoner ferromagnetic instability~\cite{stoner}. Rather convincing experimental signatures of ferromagnetic behavior have been recently observed~\cite{valtolina2017exploring}, following the pioneering experiment of Ref.~\cite{jo2009itinerant} and the theoretical analyses of Refs.~\cite{PhysRevLett.103.207201,pilati2010,trivedi,PhysRevA.79.053606,PhysRevLett.110.230401,PhysRevA.81.041602,polls,massignan2011repulsive,massignan2014polarons}.
The experimental investigation of Stoner ferromagnetism is hindered by the increase of three-body recombinations in the strongly-interacting regime of the upper branch~\cite{PhysRevLett.106.050402,ketterle2012,ketterle2012prl}.
Several mechanisms have been proposed to shift the ferromagnetic critical point to weaker interactions, including: adding shallow optical lattices~\cite{dft,pilati2014}, introducing disorder~\cite{pilati2016ferromagnetism}, tuning the mass imbalance~\cite{PhysRevA.83.053625,PhysRevLett.110.165302,fratini2014zero}, or including odd-wave interactions~\cite{Kurlov:PRA2019,jiang2016itinerant,PhysRevA.102.053301}.
Setups with reduced dimensionality have been considered as well~\cite{PhysRevA.82.043604,PhysRevB.87.184414,PhysRevA.89.023611,PhysRevA.87.060502,PhysRevLett.111.045302}, but various issues remain to be investigated, in particular for two-dimensional (2D) geometries.
The studies on three-dimensional (3D) systems found that, at large population imbalance, both attractive and repulsive Fermi gases are well described by the so-called Landau-Pomeranchuk energy functional~\cite{PhysRevLett.97.200403,pilati2010,PhysRevLett.100.030401,PhysRevLett.104.230402}. This describes quasiparticles, commonly referred to as (Fermi) polarons, that represent the building block of Landau Fermi liquids. Many of their properties have been theoretically and experimentally investigated, including their chemical potential, effective mass, and lifetime, both in two and in three dimensions~\cite{PhysRevLett.103.170402,PhysRevLett.102.230402,kohstall2012metastability,koschorreck2012attractive,PhysRevLett.122.193604}.
However, the applicability of the Landau-Pomeranchuk functional to 2D Fermi gases is still unclear.

The theoretical analysis of atomic gases in the strongly-interacting regime requires nonperturbative approaches. 
Previous computational studies for Fermi gases with short-range repulsive interactions employed, e.g., quantum Monte Carlo (QMC) simulations~\cite{PhysRevLett.103.207201,pilati2010,trivedi} and diagrammatic resummation techniques~\cite{he2014finite,he}, and they focused mostly on 3D geometries. 
2D setups have been considered for attractive interactions~\cite{Bertaina_BCSBECCrossoverTwoDimensional_2011,PhysRevA.93.023602,bauer_universal_2014,Shi_Groundstatepropertiesstrongly_2015,PhysRevA.101.033601,anderson_pressure_2015}, for repulsive gases with balanced populations~\cite{bertaina2013two}, for single polarons or for the highly-polarized regime~\cite{PhysRevA.100.023608,NgampruetikornRepulsivepolaronstwodimensional2012,SchmidtFermipolaronstwo2012,Parish_HighlypolarizedFermi_2013,VlietinckDiagrammaticMonteCarlo2014,Adlong_QuasiparticleLifetimeRepulsive_2020}, or for longer-range interactions~\cite{PhysRevB.87.184414,PhysRevA.99.043609}.
2D Fermi gases with short-range repulsive interactions at finite population imbalance require further investigations.
Beyond the aforementioned analysis on the 2D Landau-Pomeranchuk functional, the magnetic susceptibility has to be determined via accurate computational approaches. 
Furthermore, the nature of the 2D Stoner ferromagnetic instability is still unclear.

In this Article, we investigate the ground-state properties of 2D two-component Fermi gases with short-range repulsive interspecies interactions. Our computations are based on diffusion Monte Carlo (DMC) simulations~\cite{doi:10.1142/1170}. The negative-sign problem is circumvented using the fixed-node constraint, which leads to a variational upper bound for the ground-state energy.
The energy per particle is determined as a function of the interaction strength and of the population imbalance.
The role of the details of the interatomic interaction is quantified by comparing results for two model potentials, namely, the hard-disk and the soft-disk potentials.
In the large imbalance regime, we analyze the applicability of the Landau-Pomeranchuk functional for 2D repulsive polarons. 
Beyond their chemical potential at zero concentration~\cite{PhysRevA.100.023608}, we determine their effective mass and their coupling parameter.
Furthermore, from low-imbalance data, we determine the magnetic susceptibility.
These quantities allow us to estimate the critical interaction strength for the Stoner ferromagnetic instability.
The transition from a paramagnetic to a partially ferromagnetic ground state is signaled by the divergence of the susceptibility. The stability region of the fully ferromagnetic phase is identified from the polaron chemical potential.
As we discuss, while the mean-field theory predicts a direct transition from a paramagnetic to a fully ferromagnetic phase~\cite{PhysRevA.82.043604,Penna_2017,PhysRevA.90.043614}, our QMC results suggest that a partially ferromagnetic phase is stable in a narrow intermediate window of the interaction parameter.

The rest of the Article is organized as follows:
the model Hamiltonian and our computational method are described in Section~\ref{SecModel}.
Some known results from perturbative expansions are reviewed in Section~\ref{SecPerturbative}.
Our QMC results for the zero-temperature equation of state, for the polaron's properties, and the analysis of the onset of ferromagnetism are reported in Section~\ref{secResults}.
 Section~\ref{Conclusions} provides a summary of the main findings and discusses some future perspectives.


\begin{figure}[tb]
\begin{center}
\includegraphics[width=1.0\columnwidth]{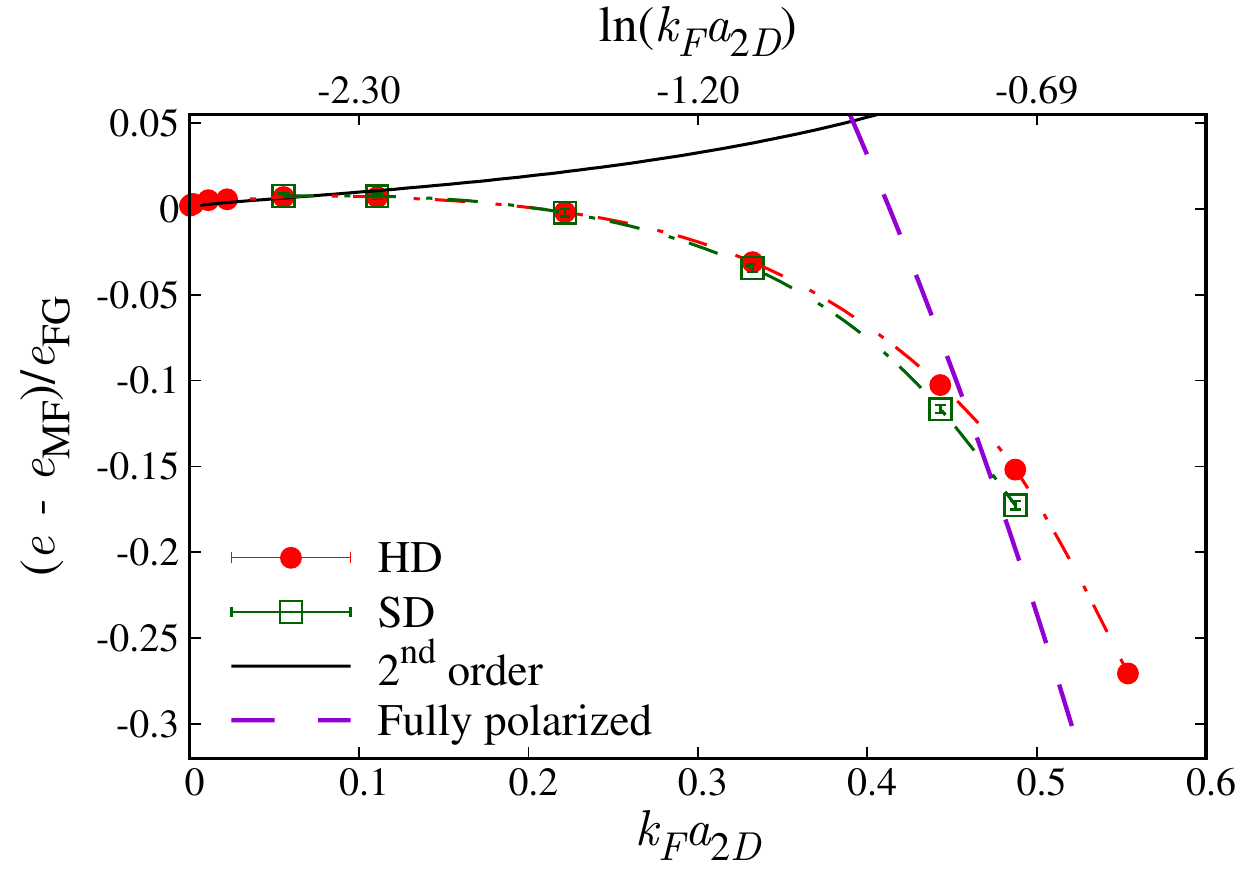}
\caption{(Color online) 
Equation of state at zero population imbalance, \emph{i.e.}, $P=0$. The dimensionless correlation energy $(e-e_{\mathrm{MF}})/e_{\mathrm{FG}}$ is plotted as a function of the dimensionless interaction parameter $k_Fa_{2D}$. $e$ is the ground-state energy per particle, $e_{\mathrm{MF}}$ is the mean-field result (see Eq.~\eqref{eqMF}), and $e_{\mathrm{FG}}$ is the energy of a balanced ideal Fermi gas. (Red) points correspond to the hard-disk intercomponent potential (HD), and (green) empty squares correspond to the soft-disk (SD) potential with range $R = 2a_{2D}$. The number of particles is $N=98$.
The continuous (black) curve indicates the second-order equation of state (see Eq.~\eqref{eq2nd}). The long-dash (purple) curve indicates the energy of the fully imbalanced ideal Fermi gas. The dot-dash curves are guides to the eye.
Here and in all figures, the errorbars are smaller than the symbol size when not visible.
}
\label{fig1}
\end{center}
\end{figure}
%
%

\section{Model and method}
\label{SecModel}
We consider 2D two-component atomic Fermi gases described by the following Hamiltonian:
\begin{equation}
H = -\sum_{\sigma=\uparrow,\downarrow}\sum_{i_\sigma=1}^{N_\sigma}\frac{\hbar^2}{2m}\nabla^2_{i_\sigma} 
+ \sum_{i_\uparrow,i_\downarrow}v(r_{i_\uparrow i_\downarrow})\; ;
\label{hamiltonian}
\end{equation}
here, $m$ is the particle mass for both components and $\hbar$ is the reduced Planck constant. The indices $i_\uparrow=1,\dots,N_\uparrow$ and $i_\downarrow=1,\dots,N_\downarrow$ label atoms of the two components, hereafter referred to as spin-up and spin-down particles. The distance between opposite-spin fermions is $r_{i_\uparrow i_\downarrow} = \left|\mathbf{r}_{i_\uparrow}-\mathbf{r}_{i_\downarrow}\right|$. The total number of fermions is $N=N_\uparrow+N_\downarrow$, and the polarization is defined as $P=(N_\uparrow-N_\downarrow)/N$.
The particles move is a square box of size $L$ with periodic boundary conditions. The total density is thus $n=N/L^2= n_\uparrow+n_\downarrow$, where the partial densities are $n_\sigma=N_\sigma/L^2$.
The latter allow defining the Fermi energies of the component $\sigma$: $E_F^\sigma = (\hbar k_F^\sigma)^2/2m$, 
where the corresponding Fermi wavevectors are $k_F^\sigma=\sqrt{4\pi n_\sigma}$. 
$v(r)$ is a short-range potential that describes the intercomponent interactions.
We consider two model potentials. The first is the hard-disk (HD) model: $v(r)=+\infty$ if $r<a_{2D}$ and zero
otherwise. The disk diameter coincides with the 2D s-wave scattering length $a_{2D}$.
The second one is the soft-disk (SD) potential: $v(r)=V_0$ if $r<R$ and zero
otherwise, where $V_0\geqslant 0$ is the potential intensity. 
In this case, the scattering length is related to the disk diameter $R$ by the relation:
\begin{equation}
a_{2D} = R\exp\left[
-\frac{1}{K_0 R}
\frac{ \mathrm{I}_0 \left( K_0 R \right)}
{\mathrm{I}_0^\prime \left( K_0 R \right) }
\right],
\end{equation}
where $K_0 = mV_0/\hbar^2$ and $\mathrm{I}_0(x)$ is the modified Bessel function of the first kind and $\mathrm{I}_0^\prime(x)$ its derivative.
In this Article, we set $R$ and $V_0$ so that $R=2a_{2D}$. This allows us to analyze the possible role played by details of the model potential beyond the s-wave scattering length. 
Notice that, due to the logarithmic dependence of the 2D scattering amplitude on energy, in the literature various definitions of the 2D scattering length have been used. 
We stick to the notation that is most natural for hard-disks -- namely, $a_{2D}$ corresponds to the disk diameter -- which was also used in Refs.~\cite{PhysRevA.71.023605,Bertaina_BCSBECCrossoverTwoDimensional_2011,bertaina2013two,PhysRevA.93.023602,Shi_Groundstatepropertiesstrongly_2015}. An alternative definition, which is often used when considering attractive interactions \cite{SchmidtFermipolaronstwo2012,koschorreck2012attractive,Adlong_QuasiparticleLifetimeRepulsive_2020}, sets the scattering length equal to $b$, where the dimer binding energy is $|\epsilon_B|=\hbar^2/mb^2$. The relation between the two definitions is $b=a_{2D}e^\gamma/2$, where $\gamma \cong 0.577$ is Euler-Mascheroni's constant, so that the coupling constant $\ln{(k_F b)}=\ln{(k_F a_{2D})}+\gamma-\ln{2}\simeq \ln{(k_F a_{2D})} - 0.12$.
In cold-atom experiments, 2D systems are created by limiting the particle motion in one direction to zero-point oscillations using a strong confining potential. The effective 2D scattering length for dilute gases is determined by solving the scattering problem in the presence of such strong confinement, integrating over virtual excitations induced by the short-range interaction~\cite{PhysRevLett.84.2551,PhysRevA.64.012706}.

To determine the ground-state properties of the Hamiltonian~\eqref{hamiltonian}, we employ the fixed-node DMC algorithm~\cite{doi:10.1142/1170}. This is designed to sample the lowest-energy wavefunction by stochastically evolving a modified Schr\"odinger equation in imaginary time. 
For stoquastic Hamiltonians, this algorithm provides unbiased estimates of the energy, provided that possible biases due to the finite time step and the finite random-walker population are reduced below the statistical uncertainty.
In order to circumvent the negative-sign problem, which affects fermionic simulations in dimensions $D>1$, the fixed-node constraint is introduced. It consists in imposing that the nodal surface of the many-body wavefunction is the same as that of a suitably chosen trial wavefunction $\psi_T$.
The predicted energies are rigorous variational upper bounds and are very close to the exact ground state energy if the nodes of $\psi_T$ are good approximations of the ground-state nodal surface.
We choose trial wavefunctions of the Jastrow-Slater type, defined as:
\begin{equation}
\psi_T({\bf r}_1,..., {\bf r}_N)= D_\uparrow(N_\uparrow) D_\downarrow(N_\downarrow) \prod_{i_\uparrow,i_\downarrow}f(r_{i_\uparrow i_\downarrow}) \;,
\label{psiT}
\end{equation}
where $D_{\uparrow(\downarrow)}$ denotes the Slater determinant of single-particle plane waves for the spin-up (spin-down) particles.
The Jastrow correlation term $f(r)$ is taken to be the solution of the s-wave radial Schr\"odinger equation describing two-particle scattering with the potential $v(r)$. The scattering energy is set so that $f^\prime(r=L/2)=0$ (for more details, see Ref.~\cite{PhysRevA.71.023605}). Since $f(r)>0$, the nodal surface is determined by the Slater determinants.

\begin{figure}[tb]
\begin{center}
\includegraphics[width=1.0\columnwidth]{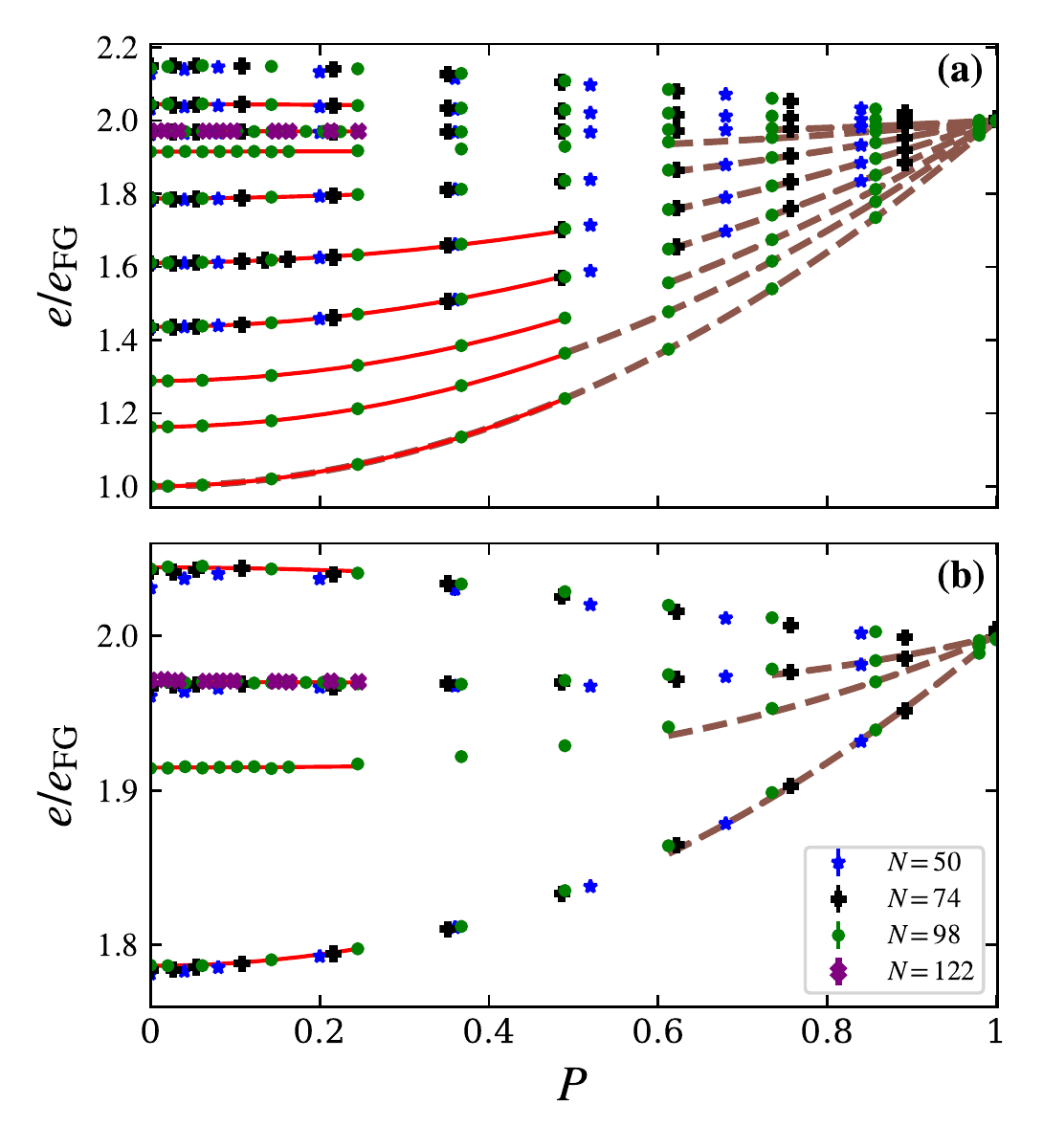}
\caption{
(Color online) 
Equation of state of imbalanced Fermi gases. The energy per particle $e/e_{\mathrm{FG}}$ is plotted as a function of the population imbalance $P$. 
$e_{\mathrm{FG}}$ is the energy of the balanced ideal Fermi gas.
Different symbols correspond to different particle numbers $N$.
Different datasets correspond to different interaction parameters $k_Fa_{2D}$, increasing from bottom to top.
Panel (a) includes data for $k_F a_{2D} \cong 0, 0.0022, 0.0332, 0.111, 
0.222, 0.332, 0.410, 0.443, 0.487, 0.554$.
Panel (b) is a zoom in the vertical axis, including only data for 
$k_F a_{2D} \cong 0.222, 0.332, 0.410, 0.443, 0.487$.
The (red) continuous curves represent the quadratic fitting functions defined in Eq.~\eqref{quadraticfit}.
The (brown) dashed curves represent the Landau-Pomeranchuk functional, namely, Eq.~\eqref{eLP}, which is applicable in the large polarization regime.
}
\label{fig2}
\end{center}
\end{figure}

Beyond the aforementioned possible biases, the QMC results might be affected by finite-size effects. To reduce them, we correct the QMC energies using the finite-size correction corresponding to noninteracting gases with the same partial densities. 
This correction is rescaled according to the particle's effective mass $m_0^*$ of the interacting system, leading to the following correction formula: $E\rightarrow E - \left[E_{\mathrm{id}}(N_\uparrow,N_\downarrow) - N_\uparrow e_{\mathrm{FG}}^\uparrow - N_\downarrow e_{\mathrm{FG}}^\downarrow \right]m/m_0^*$~\cite{PhysRevE.64.016702}, where $e_{\mathrm{FG}}^\sigma\equiv E_{\mathrm{FG}}^\sigma/N_\sigma=E_F^\sigma/2$ is the energy per particle of the ideal fully imbalanced Fermi gas of component $\sigma$, and id refers to ideal gases.
Notice that we use the same approximation for the effective mass of both components.
When extracting the magnetic susceptibility from low-imbalance data (and when displaying the QMC results in Fig.~\ref{fig2}), we use the estimate from second-order perturbation theory for balanced gases~\cite{PhysRevB.45.10135}: 
$m_0^*/m = 1 + 2 / \ln^2\left( c_0 n a_{2D}^2\right)$, where $c_0= \pi e^{2\gamma}/2\cong 4.98$. See the discussion in Section~\ref{SecPerturbative} for the choice of $c_0$.
To account for the uncertainty in the validity of the perturbative effective mass, half of the difference between the above correction and the one obtained with $m_0^*=m$ is summed in quadrature to the statistical uncertainty.
The above energy correction is implicit in the definition of the polaron chemical potential (see Section~\ref{secResults}). 
When extracting the polaron coupling parameter from high-imbalance data, we compute the energy correction with $m_0^*=m$. This choice is indeed more appropriate for the high-imbalance regime, since the interaction effects on the majority component are strongly reduced, arguably leading to an effective mass closer to the particle mass $m$.

The polaron effective mass can be determined in DMC simulations from the imaginary-time diffusion coefficient of a spin-down impurity in a spin-up Fermi sea, as~\cite{PhysRevB.59.8844,PhysRevA.92.033612}:
\begin{equation}
\label{mstarDMC}
\frac{m^*}{m}= \lim_{\tau \rightarrow \infty} \frac{m}{2\tau \hbar} \left< \left(r_{\downarrow} (\tau)- r_{\downarrow} (0)\right)^2 \right>,
\end{equation}
where $r_{\downarrow} (\tau)$ is the impurity position at imaginary time $\tau$ and the angular brackets indicate the Monte Carlo average.

%
%
%

\section{Perturbative treatment}
\label{SecPerturbative}
Whenever possible, we compare our QMC data to the corresponding perturbative results. The mean-field coupling constant 
of a dilute system is the low-energy on-shell T-matrix \cite{PhysRevB.45.10135,PhysRevLett.84.2551,Abrikosov:107441}, which reduces to the real part of the scattering amplitude. For a 2D two-component Fermi system this results in an energy-dependent coupling $\tilde{g}(E)=(4\pi \hbar^2/m)/\ln[E_a/(\mu_\uparrow+\mu_\downarrow+E)]$, where $E_a=4\hbar^2/ma_{2D}^2e^{2\gamma}$ and $\mu_{\sigma}$ is the chemical potential of the spin-$\sigma$ component. Differently from the 3D case, the zero-energy limit $E\to 0$ displays a significant density dependence, albeit logarithmic, through the chemical potentials. 
By approximating the latter with the Fermi energies $E_F^\sigma$
%
we obtain $\tilde{g}=(4\pi \hbar^2/m)/|\ln(c_0 n a_{2D}^2)|$,
where, as already mentioned in Section~\ref{SecModel}, $c_0=\pi e^{2\gamma}/2 \cong 4.98$.
Notice that only the total density $n=n_\uparrow+n_\downarrow$ appears, even in the imbalanced case, due to the linearity of the Fermi energy with the 2D density. In the literature, it is sometimes set $c_0=1$ \cite{PhysRevA.71.053605}, which is correct within logarithmic accuracy in the weak-coupling regime. However, we observe that our choice for $c_0$ allows for much more accurate perturbative expressions, when compared to the nonperturbative results, consistently with \cite{PhysRevB.45.10135,bertaina2013two,PhysRevA.100.023608}.

At the mean-field level, the energy density $\varepsilon= E/V$ in a volume $V=L^2$ is the sum of the kinetic contributions of the two components, plus an interaction term which is given by the coupling constant times the number of possible pairs, leading to:
\begin{equation}\label{eq:evMF}
 \varepsilon_{\mathrm{MF}}(n_\uparrow,n_\downarrow)=e_{\mathrm{FG}}^\uparrow n_\uparrow+e_{\mathrm{FG}}^\downarrow n_\downarrow+\tilde{g} n_\uparrow n_\downarrow\;.
\end{equation}
This standard expression is valid both for balanced and imbalanced systems~\cite{PhysRevA.71.053605}. In the balanced case ($n_\uparrow=n_\downarrow=n/2)$, it readily yields the energy per particle:
\begin{equation}\label{eqMF}
 e_{\mathrm{MF}}(n)\equiv\varepsilon_{\mathrm{MF}}(n/2,n/2)/n = e_{\mathrm{FG}}(1 + 2 g) \;,
\end{equation}
where $e_{\mathrm{FG}}=E_F/2=\pi\hbar^2 n/2m$ is the energy per particle of the ideal balanced Fermi gas, and we defined the dimensionless expansion parameter 
\begin{equation}\label{eq:coupling}
 g=\frac{1}{|\ln(c_0 n a_{2D}^2)|}.
\end{equation}
The second-order expansion in $g$ for the balanced case is also known~\cite{PhysRevB.12.125,PhysRevB.45.10135,PhysRevB.45.12419}. It reads
\begin{equation}\label{eq2nd}
e_{2^{\mathrm{nd}}}(n) = e_{\mathrm{FG}}[1 +2g+ (3-4 \ln 2 ) g^2].
\end{equation}
The coefficient of the second-order term is fixed by the choice of $c_0$.

The single-polaron chemical potential is the energy of a single spin-down impurity in a Fermi sea of the majority spin-up component. At the mean-field level, it can be derived from Eq.~\eqref{eq:evMF}, and, in dimensionless form, it reads \cite{PhysRevA.100.023608}:
\begin{equation}\label{polaronMF}
 A_{\mathrm{MF}} =L^2 \left[\varepsilon_{\mathrm{MF}}(n_\uparrow,1/L^2)-\varepsilon_{\mathrm{MF}}(n_\uparrow,0) \right]/e_{\mathrm{FG}}^{\uparrow}= 4 g\;,
\end{equation}
where we eventually took the thermodynamic limit and $g$ is calculated at $n=n_\uparrow$.

\begin{figure}[tb]
\begin{center}
\includegraphics[width=1.0\columnwidth]{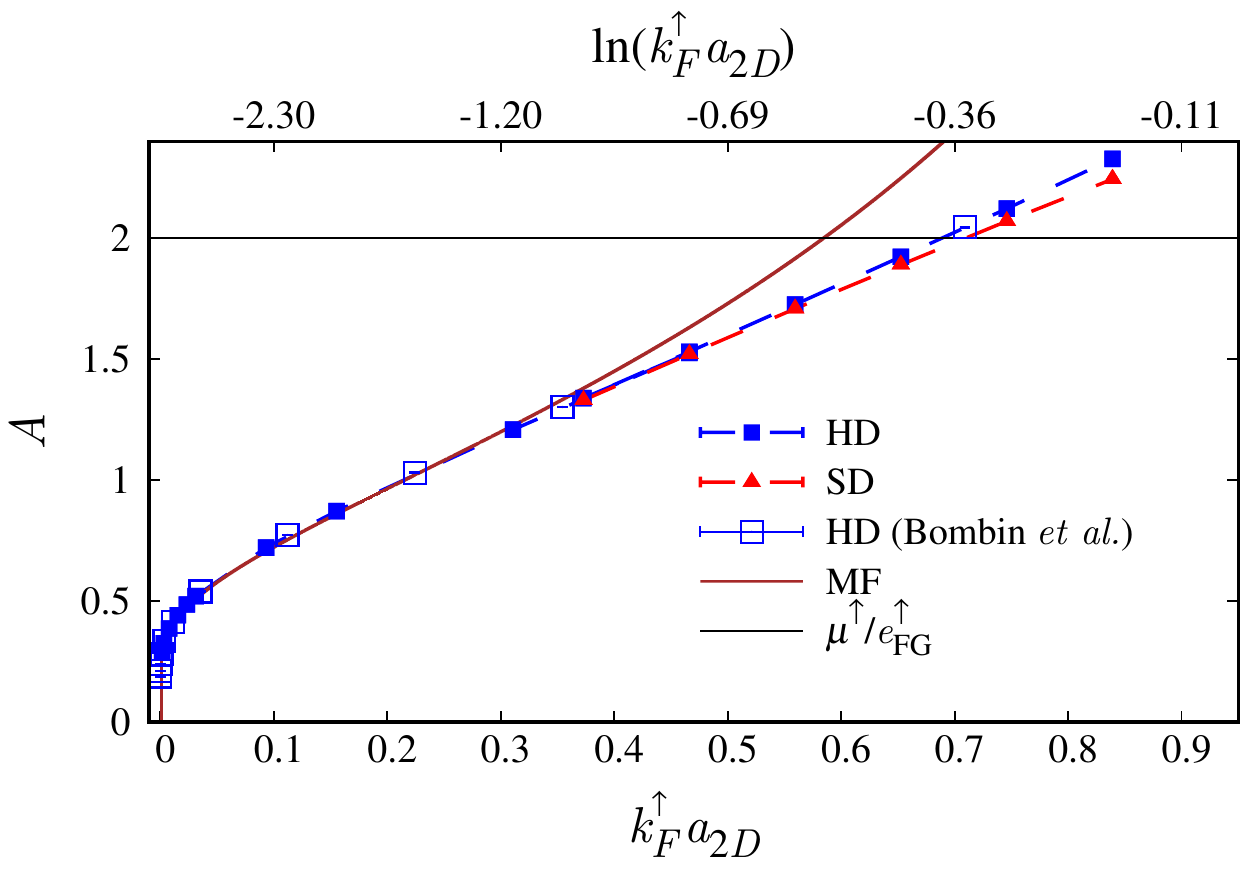}
\caption{
(Color online) 
Chemical potential at zero concentration of the repulsive polaron $A$ as a function of the interaction parameter $k_{F}^\uparrow a_{2D}$, in units of the energy per particle of the fully imbalanced ideal Fermi gas $e_{\mathrm{FG}}^{\uparrow}$.
$k_{F}^\uparrow=\sqrt{4\pi n_{\uparrow}}$ is the Fermi wavevector of the fully polarized ideal Fermi gas.
(Blue) squares correspond to the hard-disk (HD) potential. (Red) triangles correspond to the soft-disk (SD) potential. The number of majority-spin particles is $N_\uparrow = 61$.
The continuous (brown) curve represents the mean-field prediction Eq.~\eqref{polaronMF}.
The thin horizontal (black) line indicates the chemical potential of the majority component $\mu_{\uparrow}$.
The empty (blue) squares represent the results from Ref.~\cite{PhysRevA.100.023608}.
}
\label{fig3}
\end{center}
\end{figure}

\section{Results}
\label{secResults}
The zero-temperature equation of state for balanced populations (corresponding to the polarization $P=0$) has been investigated in Ref.~\cite{bertaina2013two}.
In Fig.~\ref{fig1}, our QMC results for the energy per particle $e=E/N$ are plotted as a function of the dimensionless interaction parameter $k_F a_{2D}$, where $k_F = \sqrt{2\pi n}$ is the Fermi wavevector of the balanced ideal Fermi gas.
To better visualize the interaction effects, we subtract the mean-field prediction Eq.~\eqref{eqMF}.
We also show the perturbative second-order result Eq.~\eqref{eq2nd} with the continuous black curve. Evidently, this second-order expansion is valid only for relatively weak interactions $k_{F} a_{2D} \lesssim 0.1$ \cite{bertaina2013two}.
At relatively strong interactions $k_F a_{2D} \simeq 0.45$, the energy of the balanced gas overcomes the one of the fully polarized configuration (corresponding to $P=1$). This implies that the paramagnetic phase is unstable~\cite{trivedi}. In this regime, the results for the HD and the SD potentials deviate by less than $1\%$. This indicates that the onset of ferromagnetism is essentially universal in terms of the 2D s-wave scattering length, while other details of the interaction potential play a marginal role.

\begin{figure}[tb]
\begin{center}
\includegraphics[width=1.0\columnwidth]{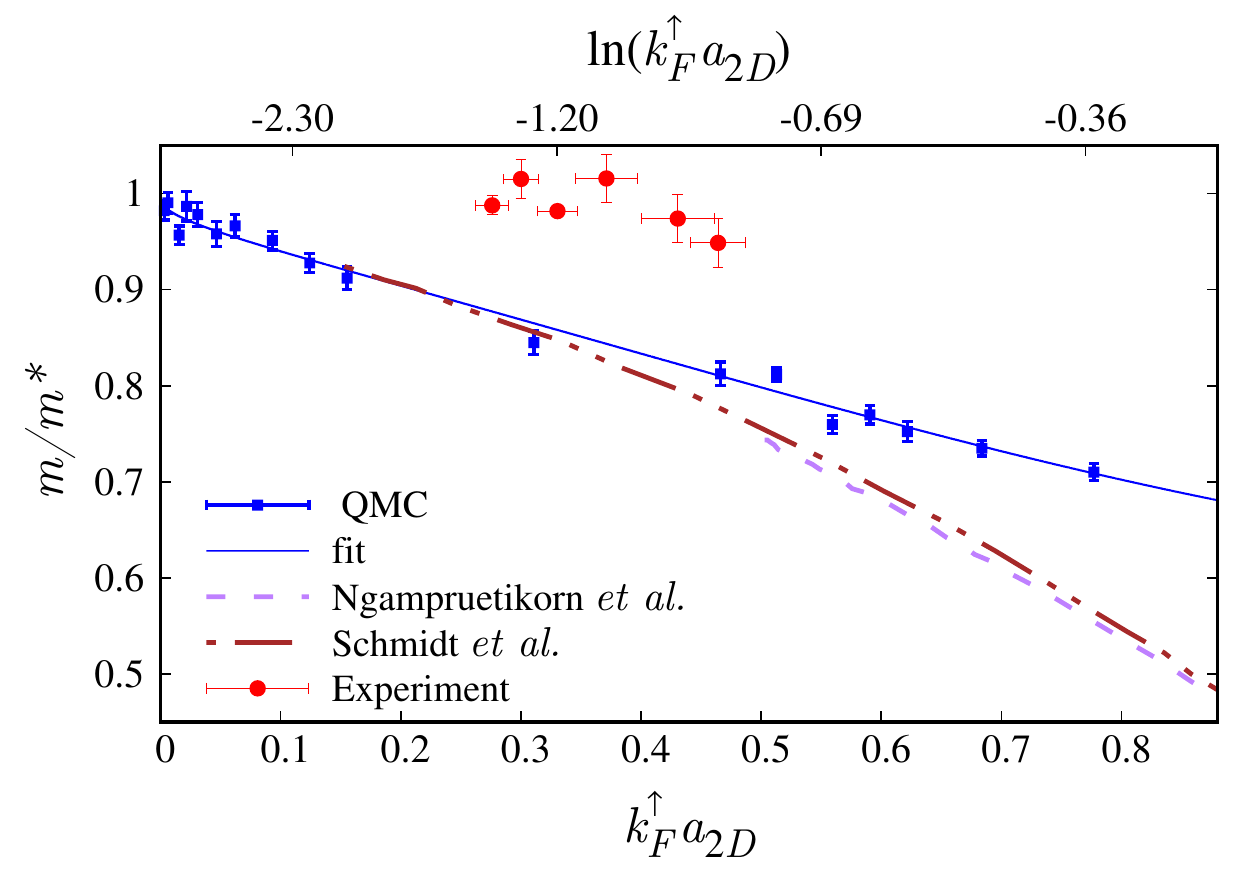}
\caption{
(Color online) 
Inverse effective mass $m/m^*$ as a function of the interaction parameter $k_{F}^\uparrow a_{2D}$. $m$ is the particle mass.
(Blue) squares represent the QMC results for the HD potential. 
The continuous (blue) line is a Pad\'e fitting function (see text).
The dashed (purple) curve and the dot-dash (brown) curve represent two previous theoretical predictions for upper-branch polarons, extracted from Refs.~\cite{NgampruetikornRepulsivepolaronstwodimensional2012} and \cite{SchmidtFermipolaronstwo2012}, respectively.
The (red) circles represent the experimental results, extracted from Ref.~\cite{koschorreck2012attractive}.
}
\label{fig4}
\end{center}
\end{figure}

Precisely locating the transition from the paramagnetic to a ferromagnetic ground-state requires simulating imbalanced components.
In Fig.~\ref{fig2}, the energy per particle $e$ is plotted as a function of the polarization $P$, for several values of the interaction parameter $k_F a_{2D}$.
One notices that the QMC results for the particle numbers $N\geqslant 74$ are in good agreement, indicating that the correction described in Section~\ref{SecModel} is adequate to suppress finite-size effects.
Interestingly, in the high-imbalance regime, corresponding to $P \simeq 1$, the equation of state is well approximated by the Landau-Pomeranchuk energy functional:
\begin{equation}
\label{eLP}
E_{\mathrm{LP}} = E_{\mathrm{FG}}^{\uparrow} \left(1 + A x +\frac{m}{m^*} x^2+ F x^2\right),
\end{equation}
where $x=N_\downarrow/N_\uparrow $ is the concentration of the minority component, here identified with the spin-down particles. The first term on the right hand side is the kinetic energy of the majority component; the second term is proportional to the dimensionless chemical potential at zero concentration of the polarons $A(k_F^\uparrow a_{2D})$; the third term represents the kinetic energy of the polaron gas, and it is fixed by the polaron effective mass $m^*(k_F^\uparrow a_{2D})$.
The fourth term represents the energy contribution due to correlations among polarons, and its magnitude is fixed by the dimensionless coupling parameter $F(k_F^\uparrow a_{2D})$. 
A quadratic scaling with the concentration, namely $F x^2$, is found to accurately describe the QMC data. If the exponent of $x$ is used as a fitting parameter, the results are compatible with the quadratic Ansatz. 
In fact, this is the same scaling of the 3D case~\cite{PhysRevLett.97.200403,PhysRevLett.104.230402}. An additional logarithmic factor might apply to the 2D case, but it would not be noticeable on the available range of concentration values.
It is worth noticing that, in two dimensions, both the third and the fourth terms scale with the second power of the concentration $x$, as opposed to the 3D case, where the former scales as $\frac{m}{m^*} x^{5/3}$~\cite{PhysRevLett.97.200403}.
In Fig.~\ref{fig2}, Eq.~\eqref{eLP} is compared to the QMC datasets for different (fixed) values of $k_F a_{2D}$, for varying $P$. The following conversion formulas are used: $k_F^\uparrow = k_F\sqrt{\frac{2}{1+x}}$, $E_{\mathrm{FG}}^{\uparrow}=2E_{\mathrm{FG}}/(1+x)^2$, and $x=\frac{1-P}{1+P}$. 
The polaron chemical potential is determined from QMC simulations as the fixed-volume energy difference $A=\left[E(N_\uparrow,1)-E(N_\uparrow,0) \right] / e_{\mathrm{FG}}^{\uparrow} $, where $E(N_\uparrow,N_\downarrow)$ is the energy of a gas with $N_{\uparrow}$ spin-up and $N_{\downarrow}$ spin-down particles. Our results are shown in Fig.~\ref{fig3}. They are compared with the QMC data from Ref.~\cite{PhysRevA.100.023608}, finding excellent agreement.
The mean-field prediction of Eq.~\eqref{polaronMF} is also shown. It appears to be accurate only in the regime $k_F^\uparrow a_{2D} \lesssim 0.3$.
When $A$ exceeds the chemical potential of the majority component $\mu_\uparrow=E_{F}^\uparrow$, a state with two fully separated domains, each hosting one component only, is thermodynamically stable~\cite{pilati2010}.
This criterion allows one to pinpoint the onset of full ferromagnetism at $k_F^\uparrow a_{2D} \simeq 0.69$, corresponding to $k_F a_{2D} \simeq 0.49$ for a gas with globally balanced populations.
Again, the small deviations between the HD and the SD results indicate the marginal role played by nonuniversal details beyond the s-wave scattering length.
If one uses the mean-field result Eq.~\eqref{polaronMF}, the condition for the stability of the fully ferromagnetic state reads $A_{\mathrm{MF}}=4g>2$. This corresponds to the critical interaction parameter $k_F a_{2D} \cong 0.413$.
The polaron effective mass $m^*$ is determined from QMC simulations with a single spin-down impurity via Eq.~\eqref{mstarDMC}. The results for the HD potential are shown in Fig.~\ref{fig4}. Previous studies reported corresponding QMC data for dipolar interactions and for soft-core potentials with large effective range~\cite{PhysRevA.100.023608,bombin2020finiterange}.
Notably, the effective mass increases up to $m^* \approx 1.35m$ in the regime where ferromagnetism occurs. In three dimensions, the corresponding effect is significantly smaller, with the effective mass reaching $m^* \approx 1.1m$~\cite{dft}. This suggests that correlation effects are more relevant in two dimensions.
In Fig.~\ref{fig4}, we compare our QMC predictions against two previous theories for the upper branch of resonant attractive interactions. The first is the Nozi\`eres-Schmitt-Rink calculation of Ref.~\cite{SchmidtFermipolaronstwo2012}, the second is the calculation based on a particle-hole variational wave-function performed in Ref.~\cite{NgampruetikornRepulsivepolaronstwodimensional2012}. While good agreement is found for moderate interactions $k_F^\uparrow \lesssim 0.4$, in the strongly-interacting regime the two previous theories appear to overestimate the effective mass compared to the QMC data. This discrepancy might be attributed to the different approximations in the compared theories, or to intrinsic differences between the repulsive model potentials adopted in our QMC simulations and the upper-branch models adopted in the two previous theories.
The experimental results of Ref.~\cite{koschorreck2012attractive} are also shown in Fig.~\ref{fig4}. They instead underestimate the polaron effective mass compared to the QMC prediction.
As argued in Ref.~\cite{koschorreck2012attractive}, the experiment might not be describable by a purely 2D model, possibly explaining this discrepancy (see also Ref.~\cite{PhysRevA.93.011603}).
%
%

We determine the polaron coupling parameter $F$ by fitting the Landau-Pomeranchuk functional Eq.~\eqref{eLP} to the HD QMC data in the high-imbalance regime, as illustrated in Fig.~\ref{fig2}.
For this fitting procedure, a predetermined parametrization of the polaron chemical potential and of the effective mass is used. 
The first is obtained from the HD data in Fig.~\ref{fig3}, which are well described by the following polynomial $A(k_F^\uparrow a_{2D})=4 g+a_A g^2 + b_A g^3 +c_A g^4$, with coefficients
$a_A = 1.52(5)$,
$b_A = -6.9(2)$, and
$c_A = 4.2(2)$.
Here, the expansion parameter $g$, defined in Eq.~\eqref{eq:coupling}, is computed at $n=n_\uparrow =k_F^{\uparrow 2}/4\pi$.
Notice that this expansion is consistent with the mean-field result Eq.~\eqref{polaronMF} in the weakly-interacting limit.
The inverse effective mass is parametrized as: $\frac{m}{m^*}(k_F^\uparrow a_{2D})=(1+a_m g^2)/(1+b_m g^2)$, with $a_m = 3.2(6)$ and $b_m =5.4(7)$ (see Fig.~\ref{fig4}).
It is interesting to observe that this parametrization is consistent with $\frac{m^*}{m}\simeq 1+2 g^2$ in the weakly-interacting limit, analogously to the perturbative result in the balanced case.
To parametrize the polaron coupling parameter, the polynomial function $F(k_F^\uparrow a_{2D})=a_F g^2 +b_F g^3$ is found to be particularly accurate. Indeed, with the optimal fitting parameters $a_F= 6.2(1)$ and $b_F =-4.6(4)$, the Landau-Pomeranchuk functional accurately describes all high-imbalance data before the ferromagnetic transition
\footnote{In Fig.~\ref{fig2}, one might notice some minuscule discrepancies. They are due to the use, in that figure, of the finite-size correction with $m_0^*$ from second-order perturbations theory. Actually, in the analysis of the Landau-Pomeranchuk functional we set $m_0^*=m$, which is more appropriate for the high-imbalance regime. See discussion in Section~\ref{SecModel}.}.
Notice that these values imply that the polaron coupling parameter is as large as, e.g., $F\approx 0.43$ for $k_F^\uparrow a_{2D} = 0.3$, indicating the relevance of interpolaron correlations.
A polynomial expansion of the polaron coupling constant $F$ was developed in Ref.~\cite{PhysRevLett.104.230402} for 3D attractive Fermi gases using the variational scheme of Ref.~\cite{PhysRevLett.98.180402}. Our result could serve as a benchmark for analogous studies for 2D systems.

\begin{figure}[tb]
\begin{center}
\includegraphics[width=1.0\columnwidth]{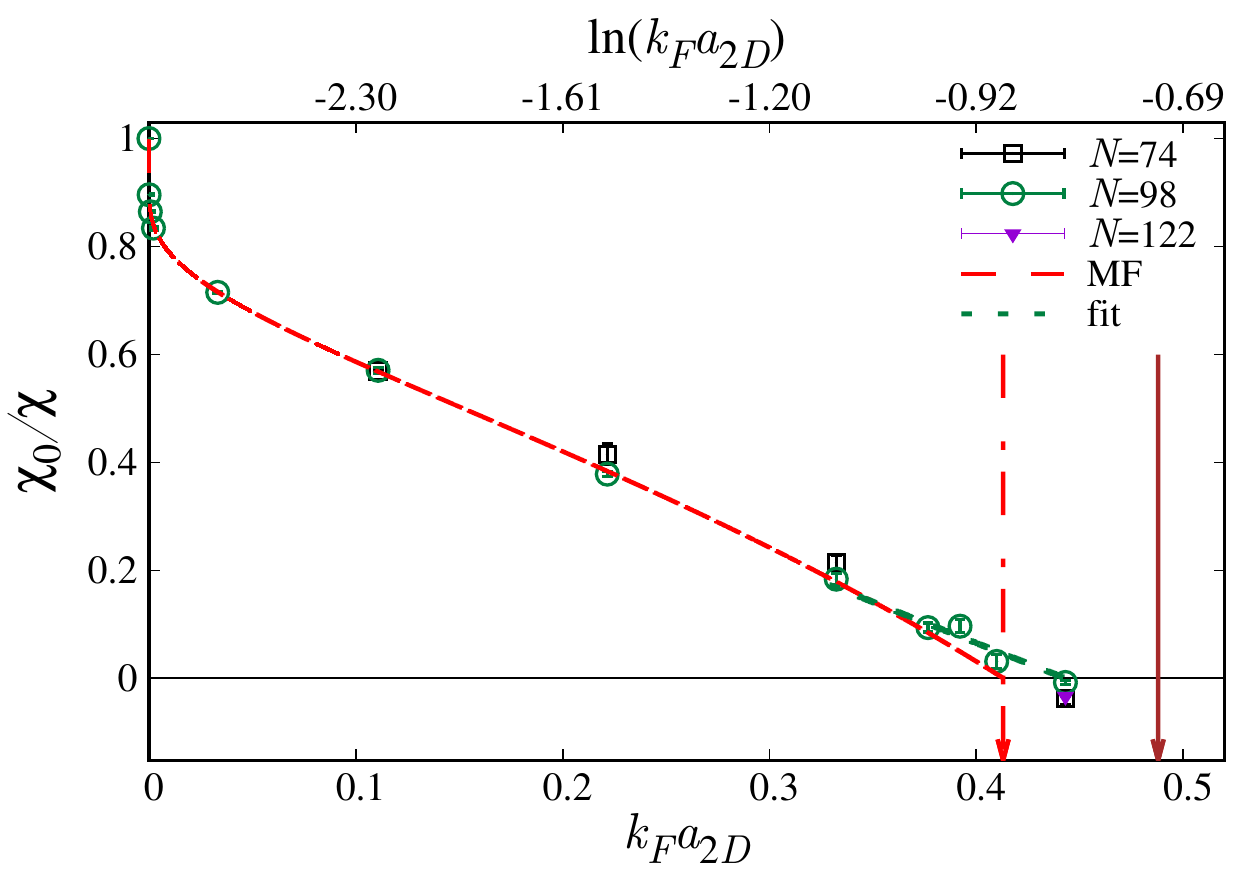}
\caption{
(Color online) 
Inverse magnetic susceptibility $\chi_0/\chi$ at $P=0$ as a function of the interaction parameter $k_{F}a_{2D}$ for the HD potential.
$\chi_0$ is the susceptibility of the ideal Fermi gas.
Different symbols correspond to different particle numbers $N$.
The long-dash (red) curve represent the mean-field (MF) prediction Eq.~\eqref{chiMF}. The short-dash (green) segment is a linear fit on $N=98$ data close to the critical point of the ferromagnetic transition.
The vertical arrows indicate the critical point for the stability of the fully ferromagnetic phase as predicted by the mean-field (MF) theory (dot-dash red arrow) and by the QMC results for the polaron chemical potential (continuous brown arrow).
}
\label{fig5}
\end{center}
\end{figure}

The stability region of the paramagnetic ground-state can be identified by analyzing the magnetic susceptibility $\chi=\left[ \frac{1}{n}\frac{\partial^2 e}{\partial P^2} \right]^{-1}$ of balanced gases.
We extract $\chi$ by fitting the low-polarization energies with quadratic functions written in the form:
\begin{equation}
\label{quadraticfit}
e(P) = e_{\mathrm{FG}} \left(a + \frac{\chi_0}{\chi} P^2\right),
\end{equation}
where $e(P)$ is the energy per particle at polarization $P$, $a$ and $\chi$ are the fitting parameters and $\chi_0=n/(2e_{\mathrm{FG}})$ is the susceptibility of the 2D balanced ideal Fermi gas.
The inverse susceptibility for the HD potential is shown in Fig.~\ref{fig5}. 
These estimates are averaged over different fitting windows, extending to different maximum polarizations from $P\simeq 0.10$ to the maximum values displayed in Fig.~\ref{fig2}.
Notably, we find good agreement with the mean-field prediction, which we derive from Eq.~\eqref{eq:evMF}:
\begin{equation}
\label{chiMF}
\chi_{\mathrm{MF}} = \chi_0 (1-2g)^{-1}\;.
\end{equation}
Small deviations occur only close to the divergence point.
This divergence signals the transition from the paramagnetic to a ferromagnetic phase. This criterion corresponds to a second-order transition. Our QMC data are indeed consistent with the second-order scenario. However, from the numerics one cannot rigorously rule out a weakly first-order transition, with two competing minima in the $e(P)$ curve -- one at $P=0$ and the other at finite $P$ -- separated by an extremely shallow maximum.
According to the MF theory, the divergence occurs when $g=1/2$. Interestingly, this critical point coincides with the MF prediction for the onset of full ferromagnetism discussed above. Therefore, the MF theory predicts a direct transition from the paramagnetic to the fully ferromagnetic phase~\cite{PhysRevA.82.043604,PhysRevA.90.043614}.
To locate the transition point from the QMC data, we perform a fit in the critical region with the scaling law $\chi \propto \left| k_F a_{2D} - k_F a_{2D}^{\mathrm{crit}} \right|^{-\gamma}$, where $\gamma=1$ is the susceptibility critical exponent for the ferromagnetic transition in metallic systems~\cite{PhysRevB.91.214407}. This fit is represented by the dashed segment in Fig.~\ref{fig5}. The best-fit parameter $k_F a_{2D}^{\mathrm{crit}} \cong 0.44$ represents an estimate of the critical interaction strength.
This value is sizably smaller than the critical point for the onset of full ferromagnetism (predicted from QMC results) discussed above, namely, $k_F a_{2D} \simeq 0.49$.
Therefore, according to the QMC data, a partially ferromagnetic phase is stable in the narrow window $0.44 \lesssim k_F a_{2D} \lesssim 0.49$.
This statement should be taken with caution. As discussed in Section~\ref{SecModel}, the QMC predictions are affected by the fixed-node constraint. It is possible that more accurate nodal surfaces based on, e.g., backflow correlations~\cite{PhysRevA.99.043609} or Pfaffian wavefunctions~\cite{PhysRevLett.96.130201,doi:10.1021/acs.jctc.0c00165}, would provide lower variational upper bounds for balanced populations, leading to a direct paramagnetic to fully-ferromagnetic transition, as predicted by the MF theory.
It is worth mentioning that beyond mean-field effects were found to open a narrow partially ferromagnetic window also in Ref.~\cite{PhysRevA.82.043604}. However, that study considered a density-independent coupling constant, which applies to quasi-2D traps in the weakly-interacting regime, where the 3D s-wave scattering length is much smaller than the cloud size in the confined direction~\cite{Bhaduri_2000}. 
It is worth emphasizing that, in two dimensions, QMC simulations locate the ferromagnetic transition at larger interaction strength compared to the mean-field prediction, as opposed to the 3D case where the mean-field theory overestimates the critical interaction strength.

The 2D Stoner ferromagnetic instability has been studied also in the 2D Hubbard model with infinite on-site repulsion~\cite{PhysRevB.83.060411}. This reference employed QMC algorithms and predicted a sharp transition from the paramagnetic to the fully ferromagnetic phase as the density increases, with a narrow intermediate region where the polarization state could not be unambiguously determined. These QMC results were described in the framework of an infinite-order phase transition~\cite{PhysRevB.16.1266}. This is characterized by the vanishing of all coefficients of the polynomial expansion of $e(P)$, leading to the flattening of the curve at the transition point. 
This scenario implies a discontinuous jump of the polarization from $P=0$ to the saturation value, as in first-order transitions, but without hysteresis. The susceptibility exponent in the paramagnetic phase is, again, $\gamma=1$~\cite{PhysRevB.16.1266}.
Similarly to the results of Ref.~\cite{PhysRevB.83.060411}, also our QMC data display a flattening of the $e(P)$ curve close to the critical point, consistently with the infinite-order transition. However, further investigations are in order to confirm the applicability of this scenario to the continuous-space 2D Fermi gas. 
%

\section{Conclusions}
\label{Conclusions}
We employed fixed-node DMC simulations to investigate the ground-state properties of 2D two-component Fermi gases with short-range repulsive intercomponent interactions. Our focus was on configurations with imbalanced populations.
Notably, we have shown that the Landau-Pomeranchuk energy functional, which was previously applied only to 3D Fermi gases, accurately describes the equation of state of 2D repulsive Fermi gases in the large imbalance regime.
Beyond the previously investigated quantities, namely, the energy of the balanced gas~\cite{bertaina2013two} and the polaron chemical potential~\cite{PhysRevA.100.023608}, we provided QMC results for the polarons' effective mass, for their coupling parameter, and for the magnetic susceptibility for balanced populations.
These are experimentally relevant properties~\cite{kohstall2012metastability}. The polaron effective mass and chemical potential have been measured in Refs.~\cite{PhysRevLett.103.170402,Navon729,PhysRevLett.103.170402,Frohlich_RadioFrequencySpectroscopyStrongly_2011,koschorreck2012attractive} for 3D or quasi-2D geometries. Ref.~\cite{koschorreck2012attractive} also observed effects due to the inter-polaron coupling for attractive interactions.
The QMC predictions for the susceptibility of 3D repulsive Fermi gases have been used in the calculations of the spin-dipole oscillation frequency~\cite{recati2011spin}, which were then employed for the experimental analysis of Stoner ferromagnetism~\cite{valtolina2017exploring}. Spin transport has also been investigated in imbalanced 2D attractive Fermi gases~\cite{Koschorreck_Universalspindynamics_2013,LuciukObservationQuantumLimitedSpin2017}.

Furthermore, we provided estimates for the critical interaction strengths where partially and fully ferromagnetic phases occur.
The mean-field theory predicts a direct paramagnetic to fully-ferromagnetic transition. Instead, our QMC results suggest that a partially ferromagnetic phase is stable in a narrow interval of the interaction parameter. 
The transition we observe is consistent with a second-order or with an infinite-order scenario.
Further investigations are in order to clarify the findings in the close vicinity of the critical point signaled by the divergence of the susceptibility, possibly with more accurate nodal surfaces. 
High-order perturbative expansions might provide useful benchmarks, as in the recent studies on 3D Fermi gases~\cite{wellenhofer2020dilute,PhysRevResearch.2.043372,wellenhofer2021effective}.
Furthermore, it is worth mentioning that more exotic ferromagnetic phases with spin textures might form in the critical region~\cite{PhysRevLett.103.207201}. 
Also, exotic pairing mechanisms~\cite{PhysRevB.48.1097} described by Pfaffian wavefunctions might be at work. We leave these analyses to future investigations.
Our results highlight interesting effects in 2D repulsive Fermi gases, and we hope that they will stimulate further experiments on this setup. The long lifetime of upper-branch polarons recently observed in the presence of orbital Feshbach resonances~\cite{PhysRevLett.122.193604}, as contrasted to the one for broad resonances~\cite{SchmidtFermipolaronstwo2012,NgampruetikornRepulsivepolaronstwodimensional2012,koschorreck2012attractive}, gives us hope that the phenomena we discussed will be experimentally realized in the near future.

All of the QMC results presented in this Article are freely available at Ref.~\cite{pilati_sebastiano_2021_4631946}.
%

\acknowledgments
S. P. acknowledges financial support from the FAR2018 project titled ``Supervised machine learning for quantum matter and computational docking'' of
the University of Camerino and from the Italian MIUR under the project PRIN2017 CEnTraL 20172H2SC4.
S. P. also acknowledges the CINECA award under the ISCRA initiative, for the availability of high performance computing resources and support.
G. O. acknowledges financial support from ANR (Grant SpiFBox) and from DIM Sirteq (Grant EML 19002465 1DFG).


%

\end{document}